\documentclass[showpacs,aps,amssymb,pra,amsmath]{revtex4}
\usepackage{float}
\usepackage{epsfig}
\usepackage{hyperref}
\usepackage{bm}
\usepackage{amsmath,graphicx}
\usepackage{subfig}
\usepackage{setspace}
\usepackage{color}

\newcommand{\beq}{\begin{equation}}
\newcommand{\eeq}{\end{equation}}
\newcommand{\bea}{\begin{eqnarray}}
\newcommand{\eea}{\end{eqnarray}}

\def\tq{{\tilde{q}}}
\def\gt{{\tilde{\gamma}}}

\def\bk{{\bf k}}

\def\bq{{\bf q}}

\def\br{{\bf r}}

\begin{document}
\title{Ordered phases in a bilayer system of dipolar fermions}
\author{B. P. van Zyl, and W. Ferguson}
\affiliation{Department of Physics, St. Francis Xavier University, Antigonish, Nova Scotia, Canada B2G 2W5}
\date{\today}

\begin{abstract}
The liquid-to-ordered phase transition in a bilayer system of fermions is studied within the context
of a recently proposed density-functional theory [Phys. Rev. A {\bf 92}, 023614 (2015)].   In each two-dimensional layer, the fermions interact via a repulsive, isotropic dipolar interaction.  
The presence of a second layer introduces an attractive {\em interlayer} interaction, 
thereby allowing for inhomogeneous 
density phases which would otherwise be energetically unfavourable.  For any fixed layer separation, we find an instability to a
commensurate one-dimensional stripe phase in each layer, which always precedes the formation of a triangular Wigner crystal.  However, at a certain {\em fixed} coupling, 
tuning the separation can lead to the system favoring a
commensurate triangular Wigner crystal, or one-dimensional stripe phase, completely bypassing the Fermi liquid state.  While other crystalline symmetries, with energies lower than the liquid phase can be found, they are never allowed to
form owing to their high energetic cost relative to the triangular Wigner crystal and stripe phase.  

\end{abstract}
\maketitle
\section{Introduction}\label{intro}
In a recent paper~\cite{vanzyl1_2015}, a density-functional theory (DFT)~\cite{DFT} was developed for investigating
the zero-temperature liquid-to-ordered phase transition in a single-component, two-dimensional (2D) dipolar Fermi
gas (dFG).  The model considered
is one in which all of the dipoles are aligned perpendicular to the 2D $xy$-plane, thereby rendering the dipole-dipole
interaction isotropic and repulsive.  The central finding of this study is that the single-layer dFG 
spontaneously breaks rotational invariance above some critical coupling to a one-dimensional
stripe phase (1DSP), followed by a transition to a triangular Wigner crystal (TWC) at slightly higher coupling.
The DFT prediction that the 1DSP precedes the formation of a 2D  TWC is supported by
earlier calculations utilizing the random-phase (RPA)\cite{Sun10,Yamaguchi10,zinner}, STLS~\cite{Parish12}, 
and conserving Hatree-Fock (HF) approximations~\cite{Sieberer11,Block12,Babadi11,block2014}.
However, recent variational and quantum Monte-Carlo (QMC) calculations~\cite{babadi2013,Mateeva12,abed12} suggest
that the TWC phase precedes the 1DSP with a coupling strength which is an order of magnitude 
larger than
any of the values predicted in
Refs.~\cite{vanzyl1_2015,Lin10,Cremon10,Sun10,Yamaguchi10,zinner,Parish12,Sieberer11,Block12,Babadi11,block2014}.
The controversy generated by the single-layer dFG studies suggests that it should also be interesting to consider
the natural extension of this system to the case of two (or more) 2D layers.  

To date, 
there are relatively few articles in the literature dealing with
the density instabilities in multilayer systems.  In Refs.~\cite{zinner,Block12,Babadi11}, the RPA~\cite{zinner} and
conserving HF~\cite{Block12,Babadi11} theoretical formulations were extended to the case of two or more layers, with a focus on investigating
the instability towards a stripe phase in each layer.  In these studies, when the intralayer dipolar interaction
is purely repulsive,
a transition to a 1DSP is found, with the stripes in each layer being ``in-phase'' (i.e., aligned with one another).  
An application of the STLS formalism to the multilayer case has also recently been provided in Refs.~\cite{marchetti1,marchetti2}, where an ``in-phase'' 1DSP was found, along with the transition 
being shifted to lower coupling compared
to the single-layer geometry.  The consensus therefore appears to be that for repulsive intralayer dipolar interactions,
a variation of the layer densitiy and/or  separation, leads to the bilayer 2D dFG
undergoing a transtion to a 1DSP,  
such that the stripes in each layer are commensurate, or ``in-phase''.   It is
worthwhile pointing out that QMC
calculations analogous to  the single-layer dFG  have yet to be performed for a multilayer configuration.

In this paper, we propose to extend the DFT of Ref.~\cite{vanzyl1_2015} to a bilayer 2D dFG, and investigate whether it 
predicts any additional density instabilities away from the uniform liquid state 
not present in the single-layer scenario.  The motivation for this study is manifold.  First, as mentioned above, there
are a limited number of papers focusing on the quantum phase transition in a multilayer dFG, 
and of these studies, none have considered the possibility of a transition to a Wigner crystal.  
In addition, there is a complete absence of any application of DFT to the 
investigation of density instabilities in the bilayer dFG; 
filling this void will allow for useful comparisons with results obtained from other theoretical approaches~\cite{zinner,Block12,Babadi11,marchetti2,marchetti1}, 
along with providing relevant predictions to be tested by any future QMC
calculations on bilayer dipolar Fermi systems.  Moreover, the mathematical formulation of  DFT is exceedingly simple, and computationally efficient, meaning
that we are able to investigate a large set of system configurations with little additional effort.
Finally,  we are encouraged by the 
early work of Goldoni and Peeters~\cite{peeters}, who implemented a DFT to examine the crystalline phases
in a bilayer {\em electron} system.  The result of their investigation was that the preferred 
ordered state in each layer is not necessarily the expected TWC.  Specifically, depending on the layer separation and electronic density, the crystal phases
could be square, rectangular, rhombic, or triangular.  In fact, 
 owing to the repulsive intralayer and interlayer Coulomb interactions, the crystalline structures in each
2D slab were found to be staggered relative to each other, so as to minimize the electrostatic 
energy of the system.  Whether the
bilayer 2D dFG possesses crystalline phases analogous to the bilayer 2D electron gas is certainly of interest, particularly in view of the fact that DFT
predicts a 1DSP, rather than a TWC phase, as the preferred ordered state in the single-layer geometry.  

To proceed, we organize the rest of our paper as follows.
In Sec.~\ref{DFTBL},  we present the DFT appropriate to
the bilayer 2D dFG, along with providing  simple analytical results for examining the transition from the liquid-to-ordered state.  
Section~\ref{phase_diagram} implements the results of Sec.~\ref{DFTBL} to generate the phase diagram for the
 bilayer system.  The
paper closes in Sec.~\ref{closing} with our conclusions and suggestions for future work.

\section{DFT  For the bilayer 2D dFG}\label{DFTBL}
Let us briefly review some of the relevant mathematical and theoretical concepts necessary for the extension of the single-layer DFT~\cite{vanzyl1_2015} to a bilayer geometry. 
We recall that for a strictly 2D dFG with all of the moments oriented along the $z$-axis (see the inset to Fig.~\ref{fig1}),
the dipole-dipole interaction is isotropic and repulsive, viz.,

\beq\label{Vdd}
V_{\rm dd}(\br - \br') = \frac{C_{\rm dd}}{4\pi |\br - \br'|^3}~,
\eeq
where $C_{\rm dd}=\mu_0 d^2$, or $C_{\rm dd}=p^2/\epsilon_0$,  for magnetic (with magnetic moment $d$) or electric dipolar atoms (with electric dipole moment $p$), respectively.  The vectors
$\br$ and $\br'$ are the coordinates of two dipoles in the $xy$-plane.   The natural length and energy scales of the
system are given by 
$r_0 = M C_{\rm dd}/(4 \pi \hbar^2)$ and
$\hbar^2/Mr_0^2$, respectively, with $M$ being the mass of an atom.  Introducing the 2D Fermi wavevector, $k_F=\sqrt{4 \pi \rho_0}$, we may define a dimensionless coupling constant
(equivalently, density), viz., $\lambda_0 = k_F r_0$, where $\rho_0$ is the areal density of the uniform system.  For a weakly inhomogeneous system, it is reasonable to introduce the local-density
approximation (LDA)~\cite{DFT}, which amounts to $\lambda_0 \to \lambda(\br)$, and $k_F \to k_F(\br)$; that is, in the LDA, the uniform density, $\rho_0$, is replaced locally by a spatially varying density, $\rho(\br)$.

As discussed in detail in Ref.~\cite{vanzyl1_2015}, the total energy functional for a single-layer (SL) 2D dFG (within the LDA) may be written as
\bea\label{EtotLDA}
E_{\rm SL}[\lambda(\br)] &=& \frac{1}{16\pi} \frac{\hbar^2}{M r_0^4}\int d^2r~\lambda(\br)^4 +  \frac{\lambda_{\rm vW}}{8\pi} \frac{\hbar^2}{M r_0^2}\int d^2r~ |\nabla \lambda(\br)|^2+
\frac{8}{45 \pi^2}\frac{\hbar^2}{M r_0^4}\int d^2r~ \lambda(\br)^5\nonumber \\
& -& \frac{C_{\rm dd}}{4 (4 \pi)^2 r_0^4}\int d^2r~ \lambda(\br)^2 \int d^2r' \int \frac{d^2k}{(2\pi)^2}~ k e^{-i\bk \cdot (\br-\br')}
\lambda(\br')^2\nonumber \\
&-&\frac{1}{32\pi}\frac{\hbar^2}{M r_0^4} \int d^2r~\lambda(\br)^6\ln \left( 1 + \frac{1}{a\sqrt{\lambda(\br)} + b \lambda(\br) + c \lambda(\br)^{\frac{3}{2}}}\right)~,
\eea
where $a = 1.1958$, $b=1.1017$, and $c=-0.0100$~\cite{abed12}.
In Eq.~\eqref{EtotLDA}, the first term is the Thomas-Fermi kinetic energy functional for a noninteracting Fermi system, 
the second term is the von Weizs{\"a}cker-like (vW) gradient correction to the kinetic energy~\cite{vanzyl_pisarski}, the third and fourth terms
correspond to the LDA for the {\em total} dipole-dipole HF interaction energy (i.e., they {\em are not} separately the direct and exchange terms, respectively)~\cite{fang,vanzyl_pisarski}, 
while the last term corresponds to the correlation energy~\cite{abed12}.   Note that for a uniform system, $\rho(\br) \to \rho_0$, and the second and fourth
terms in Eq.~\eqref{EtotLDA} vanish.

The total energy functional, $E_{\rm SL}$, can then be used to investigate instabilities away from the Fermi liquid state by considering the quantity~\cite{ghosh95}
\beq\label{deltaE}
\Delta \varepsilon_{\rm SL} = \varepsilon_{\rm inhomo} - \varepsilon_{\rm uniform} = \frac{E_{\rm SL}[\lambda(\br)] - E_{\rm SL}[\lambda_0]}{\int d^2r~\rho_0}\left(\frac{\hbar^2}{Mr_0^2}\right)^{-1}~,
\eeq
which represents the {\em dimensionless} difference in energy (per particle) between the inhomogeneous and 
uniform phases.  The results of applying Eq.~\eqref{deltaE} to various inhomogeneous densities are
presented in Ref.~\cite{vanzyl1_2015}, where it was
found that the nonlocal contribution to the HF interaction energy, viz., the fourth term in Eq.~\eqref{EtotLDA}, is crucial for the onset of a phase transition from the liquid to 1DSP.  In other words,
without the nonlocal term, the Fermi liquid state is always stable toward any ordered phase in the single-layer 2D dipolar Fermi gas.

The essential ingredient for extending Eq.~\eqref{EtotLDA} to the case of a bilayer system is  the interlayer interaction energy between dipoles residing in different 2D slabs, viz.,

\beq\label{EI}
E_{\rm I}[\rho] = \int d^2r \int d^2r' \rho_2(\br') V_{\rm I}(\br-\br';\gamma)\rho_1(\br)~,
\eeq
where $\rho_1(\br)$ and $\rho_2(\br')$, and $\br$ and $\br'$, are the areal densities and spatial coordinates in layers 1 and 2, respectively (see the inset to Fig.~\ref{fig1}).  
The interlayer interaction potential, $V_{\rm I}$, reads
\beq\label{VI}
V_{\rm I}(\br - \br';\gamma) = \frac{C_{\rm dd}}{4 \pi} \frac{|\br - \br'|^2 - 2 \gamma^2}{(|\br - \br'|^2 + \gamma^2)^{5/2}}~,
\eeq
which is attractive for $|\br - \br'|^2 < 2\gamma^2$ and repulsive for $|\br - \br'|^2 > 2\gamma^2$, where $\gamma$ is the separation between the two layers.
Equation~\eqref{EI} is most easily evaluated by considering the Fourier transform of Eq.~\eqref{VI}, viz.,
\bea\label{VIFT}
V_{\rm I}(\br-\br';\gamma)& =& \frac{1}{(2\pi)^2} \int d^2k~\tilde{V}_{\rm I}(\bk;\gamma) e^{-i\bk \cdot(\br - \br')}\nonumber \\
&=& -\frac{C_{\rm dd}}{2} \frac{1}{(2\pi)^2} \int d^2k~ k e^{-k \gamma} e^{-i\bk \cdot(\br - \br')}~,
\eea
where $\bk$ is the variable conjugate to $\br - \br'$, and $k=|\bk|$.  The interlayer interaction energy then reads
\beq\label{EIFT}
E_{\rm I}[\rho] =- \frac{C_{\rm dd}}{2}\int d^2r~ \rho_1(\br) \int d^2r' \int \frac{d^2k}{(2\pi)^2}~ ke^{-k \gamma} e^{-i\bk \cdot (\br-\br')}
\rho_2(\br')~,
\eeq
which bears a striking resemblance to the nonlocal HF interaction energy, $E^{(2)}_{\rm dd}$, for the {\em single-layer} system.  Indeed, recall that~\cite{fang,vanzyl_pisarski}
\bea\label{hfE2}
E_{\rm dd}^{(2)}[\rho] =- \frac{C_{\rm dd}}{4}\int d^2r~ \rho(\br) \int d^2r' \int \frac{d^2k}{(2\pi)^2}~ k e^{-i\bk \cdot (\br-\br')} \rho(\br')~.
\eea
 Equation~\eqref{hfE2} corresponds exactly to the fourth term in Eq.~\eqref{EtotLDA} provided we use the definition $\lambda(\br) = \sqrt{4 \pi \rho(\br)}$. 
 Now, comparing Eq.~\eqref{EIFT} to Eq.~\eqref{hfE2}, we observe that $E^{(2)}_{\rm dd}$, for the
{\em single-layer} system, can evidently be obtained (aside from a factor of 2) 
by simply evaluating the direct interaction between two 2D layers separated by a distance $\gamma$, and then taking the $\gamma \to 0$ limit.  In order to understand this result, let us re-write 
Eq.~\eqref{EI} as
\beq
E_{\rm I}[\rho] = \int d^2 r~V_2(\br) \rho_1(\br)~,
\eeq
where 
\beq
V_2(\br) = \int d^2r' \rho_2(\br') V_{\rm I}(\br-\br';\gamma)
\eeq
can be thought of as the potential produced by $\rho_2(\br)$.  We see that if $\rho_2(\br) = \rho_0$
is uniform, then $V_2(\br)=0$, meaning that $E_{\rm I}=0$ regardless of the form of $\rho_1(\br)$.  Thus, it is only the deviations from the uniform density, $\rho_0$,
which contribute to the interlayer interaction.  Now, as we take $\gamma \to 0$, one can
think of $E_{\rm I}$ as being the interaction energy between density fluctuations of two independent gases
superposed on each other.  With these comments in mind, an interesting interpretation of
$E^{(2)}_{\rm dd}$ develops; namely, $E^{(2)}_{\rm dd}$ represents the classical 
{\em self-interaction} energy between density fluctuations in the gas of  a {\em single-layer} 2D dFG.  
Viewing
$E^{(2)}_{\rm dd}$ in this way also naturally explains the factor of 2 mentioned
above, in that we now have the relationship,
\beq\label{connect}
E^{(2)}_{\rm dd} = \frac{1}{2} \lim_{\gamma\to 0} E_{\rm I}~,
\eeq
where the $\gamma \to 0$ limit is taken after the integration, since Eq.~\eqref{EI} 
is not convergent for $\gamma = 0$.  In this sense, for a single 2D dFG, $\gamma$ is simply a parameter
which causes the divergent bare dipolar self-interaction energy integral to converge, after which one
can  set $\gamma=0$. 
The connection displayed in Eq.~\eqref{connect} does not appear to have been noticed before, an
 provides an alternative route to 
obtaining the interaction energy functional (within the HF approximation) 
for a weakly modulated 2D dFG.  
In what follows, we shall make use of Eq.~\eqref{connect} to immediately construct
the interlayer energy, $E_{\rm I}$, without having to perform any additional calculations.

We are now in a position to write the total energy functional for the bilayer (BL) 2D dFG, namely,
\bea\label{EtotBL}
E_{\rm BL}[\lambda(\br)] &=& E_{\rm SL}[\lambda(\br)] + E_{\rm I}[\lambda(\br)]
\nonumber \\
&=&  \frac{1}{16\pi} \frac{\hbar^2}{M r_0^4}\int d^2r~\lambda(\br)^4 +  \frac{\lambda_{\rm vW}}{8\pi} \frac{\hbar^2}{M r_0^2}\int d^2r~ |\nabla \lambda(\br)|^2+
\frac{8}{45 \pi^2}\frac{\hbar^2}{M r_0^4}\int d^2r~ \lambda(\br)^5\nonumber \\
& -& \frac{C_{\rm dd}}{4 (4 \pi)^2 r_0^4}\int d^2r~ \lambda(\br)^2 \int d^2r' \int \frac{d^2k}{(2\pi)^2}~ k e^{-i\bk \cdot (\br-\br')}
\lambda(\br')^2\nonumber \\
&-&\frac{1}{32\pi}\frac{\hbar^2}{M r_0^4} \int d^2r~\lambda(\br)^6\ln \left( 1 + \frac{1}{a\sqrt{\lambda(\br)} + b \lambda(\br) + c \lambda(\br)^{\frac{3}{2}}}\right)\nonumber \\
& -& \frac{C_{\rm dd}}{2 (4 \pi)^2 r_0^4}\int d^2r~ \lambda_1(\br)^2 \int d^2r' \int \frac{d^2k}{(2\pi)^2}~ k e^{-k \gamma}  e^{-i\bk \cdot (\br-\br')}
\lambda_2(\br')^2~.
\eea
A few comments are in order at this point.  First, note that in Eq.~\eqref{EtotBL}, we have ignored
the thickness of the layers, and have prohibited the possibility of  tunneling between the layers.  
In addition, while $E_{\rm BL}[\lambda]$ does contain the intralayer correlations, we have not taken into account the
interlayer correlations; in the absence of any QMC calculations for the bilayer geometry, this is an unavoidable approximation.  Nevertheless, we expect that omitting 
the interlayer correlations should not qualitatively affect the outcome of this work, since it is the intralayer correlations which largely determine the transition from the liquid-to-ordered 
state~\cite{vanzyl1_2015,peeters}.

Following the methodology of the single-layer case, we will  now investigate if the bilayer 2D dFG has 
a propensity to possess ordered states by employing  Eq.~\eqref{deltaE}, but now with $E_{\rm BL}[\lambda]$ replacing $E_{\rm SL}[\lambda]$,
\bea\label{deltaEBL}
\Delta \varepsilon_{\rm BL} &=&  \frac{E_{\rm BL}[\lambda(\br)] - E_{\rm BL}[\lambda_0]}{\int d^2r~\rho_0}\left(\frac{\hbar^2}{Mr_0^2}\right)^{-1}
\nonumber \\
&=&   \frac{E_{\rm SL}[\lambda(\br)] - E_{\rm SL}[\lambda_0]}{\int d^2r~\rho_0}\left(\frac{\hbar^2}{Mr_0^2}\right)^{-1} +  \frac{E_{\rm I}[\lambda(\br)]}{\int d^2r~\rho_0}\left(\frac{\hbar^2}{Mr_0^2}\right)^{-1}\nonumber\\
&=& \Delta \varepsilon_{\rm SL} + \Delta \varepsilon_{\rm I}~,
\eea
where we have made use of the fact that $E_{\rm I}[\lambda_0]=0$. 
We will provide details for density modulations corresponding to the 1DSP and  TWC 
 in each layer, noting that other crystalline symmetries follow exactly the same analysis.


\subsection{Density modulation for a 1DSP}

The simplest {\em ansatz} corresponding to a 1D spatial modulation, with wavevector $\bq$, is given by
\beq\label{denmod}
\rho(\br) = \rho_0(1 + \alpha \cos(\bq\cdot \br))~,
\eeq
where $\alpha \ll 1$ represents the small amplitude variation about the uniform density, $\rho_0$.   Equation~\eqref{denmod} can also be cast in terms of the dimensionless quantity, $\lambda(\br)$, viz.,
\beq\label{lambdamod}
\lambda(\br) = \lambda_0(1+\alpha\cos(\bq \cdot \br))^{\frac{1}{2}}~.
\eeq
In the case of two layers, we should also take into account the possibility that the 
modulated densities are not commensurate,
namely, that they may be offset relative to each other by some phase-shift, $\phi$, viz.,
\beq\label{den1}
\rho_1(\br) = \rho_0(1 + \alpha \cos(\bq\cdot \br))~,
\eeq
\beq\label{den2}
\rho_2(\br') = \rho_0(1 + \alpha \cos(\bq\cdot \br'+\phi))~.
\eeq
Using Eqs.~\eqref{den1} and \eqref{den2}  in Eq.~\eqref{EIFT}, we obtain for $\Delta\varepsilon_{\rm I}$,
\bea\label{deltaEI}
\Delta\varepsilon_{\rm I} = -\frac{1}{4}\alpha^2\tilde{q} e^{-\tilde{\gamma}\tilde{q}}\lambda_0^2~\cos(\phi)~,
\eea
where we have introduced the dimensionless quantities $\tilde{q} = q r_0$ and $\tilde{\gamma}=\gamma/r_0$.  Note that $\Delta \varepsilon_{\rm I}$ vanishes when $\gamma \to \infty$, or when the stripe
phases in the layers are shifted by $\phi=\pi/2$ ${\rm mod}(2\pi)$.  Regardless, it is clear that Eq.~\eqref{deltaEI}
favours the situation in which the stripes in each layer are commensurate, viz., $\phi=0$.  Our prediction that the
1DSP in each layer are ``in-phase'' is supported by the other calculations~\cite{zinner,Block12,Babadi11,marchetti2,marchetti1}, although our determination is obtained via a simple
analytical expression, Eq.~\eqref{deltaEI}, rather than relying on involved numerical computations.

Coming back to Eq.~\eqref{deltaEBL}, we now make use of the fact that $\alpha \ll 1$, so that a perturbative 
approach is sensible, meaning that we need only consider shifts
to ${\cal O}(\alpha^2)$.  It is evident from Eq.~\eqref{deltaEI} that the shift from the interlayer interaction is already ${\cal O}(\alpha^2)$.  Utilizing the result
for $\Delta \varepsilon_{\rm SL}$ from Ref.~\cite{vanzyl1_2015}, 

\beq\label{deltaESL}
\frac{\Delta {\varepsilon}_{\rm SL}}{\alpha^2} = \frac{1}{8}\lambda_0^2 +\lambda_{\rm vW} \frac{\tq^2}{16} +
\left[ \frac{2}{3\pi} - \frac{1}{8} \frac{\tq}{\lambda_0} \right]      \lambda_0^3
 -\frac{1}{8}\lambda_0^4 \left[ \frac{3}{2} \ln [f_0] + \frac{11}{16}\lambda_0 A + \frac{1}{16} \lambda_0^2 B\right]~,
\eeq
where
\bea
f(\lambda) &=& \left( 1 + \frac{1}{a\sqrt{\lambda} + b \lambda + c \lambda^{\frac{3}{2}}}\right)~,
\\
f_0 &\equiv& f(\lambda_0),~~~
f'_0 \equiv \left. \frac{df}{d\lambda}\right |_{\lambda=\lambda_0},~~~
f''_0 \equiv \left. \frac{d^2f}{d\lambda^2}\right |_{\lambda=\lambda_0}~,\\
A&=& \frac{f_0'}{f_0}~,\\
B&=&\frac{f''_0f_0-f_0'^2}{f_0^2}~,
\eea
we finally obtain our desired result:
\bea\label{deltaEBL_2}
\frac{\Delta \varepsilon^{\rm 1DSP}_{\rm BL}}{\alpha^2} &=&  \frac{1}{8}\lambda_0^2 +\lambda_{\rm vW} \frac{\tq^2}{16} +
\left[ \frac{2}{3\pi} - \frac{1}{8} \frac{\tq}{\lambda_0} \right]      \lambda_0^3
 -\frac{1}{8}\lambda_0^4 \left[ \frac{3}{2} \ln [f_0] + \frac{11}{16}\lambda_0 A + \frac{1}{16} \lambda_0^2 B\right] -\frac{1}{4}\tilde{q} e^{-\tilde{\gamma}\tilde{q}}\lambda_0^2~.
\eea
Recall that all energies are scaled by $\hbar^2/Mr_0^2$.
Equation~\eqref{deltaEBL_2} illustrates that the interlayer interaction will actually enhance the system's tendency to undergo a transition to a 1DSP  because it contributes
a negative component to the energy shift, $\Delta \varepsilon^{\rm 1DSP}_{\rm BL}$.  Indeed,  it was found in Ref.~\cite{vanzyl1_2015} that a negative contribution to the energy shift  
(arising from the nonlocal HF interaction, Eq.~\eqref{hfE2}),
was essential for the single-layer system to undergo a phase transition.  The additional term coming from 
Eq.~\eqref{deltaEI} will clearly serve to lower the system's energy in favour of an
inhomogeneous state in the bilayer configuration, and therefore {\em lower} the coupling, $\lambda_0$, required for the onset of the
phase transition.


\subsection{Denstiy modulation for a triangular Wigner crystal}
In order to find the expression for $\Delta \varepsilon_{\rm BL}$ corresponding to a triangular TWC, we use the densities~\cite{vanzyl1_2015,note1},
\beq\label{denmodWC}
\rho(\br)=\rho(x,y) =\rho_0 \left[\sqrt{1-\frac{3}{2}\alpha^2}+ \alpha \cos\left(q x\right)  + 2 \alpha \cos\left(\frac{q}{2}x\right) \cos\left(\frac{\sqrt{3}}{2}q y\right)   \right]^2~,
\eeq
\beq\label{lambdaWC}
\lambda(\br)=\lambda(x,y) =\lambda_0  \left[\sqrt{1-\frac{3}{2}\alpha^2}+ \alpha \cos\left(q x\right)  + 2 \alpha \cos\left(\frac{q}{2}x\right) \cos\left(\frac{\sqrt{3}}{2}q y\right)  \right]~,
\eeq
in each layer, with the understanding the the spatial coordinates are specific to a given layer.  
In view of  our findings from the 1DSP case, we do not need to consider a scenario in which the TWC 
lattices in the two layers are staggered in any way
 (it is in this sense that we mean the TWCs in each layer are commensurate).
  This is in contrast to the Coulomb
bilayer problem, where the lattices in each layer must be incommensurate in order to minimize the total energy of the system~\cite{peeters}.  

From Ref.~\cite{vanzyl1_2015} we have for the single-layer
\beq\label{deltaE3WC}
\frac{\Delta {\varepsilon}_{\rm SL}}{\alpha^2} = \frac{3}{2}\lambda_0^2 +\lambda_{\rm vW} \frac{3 \tq^2}{4} +
\left[ \frac{8}{\pi} - \frac{3}{2} \frac{\tq}{\lambda_0} \right]      \lambda_0^3
 -\frac{1}{4}\lambda_0^4 \left[ 9 \ln [f_0] + \frac{33}{8}\lambda_0 A + \frac{3}{8} \lambda_0^2 B\right]~.
\eeq
As previously advertised, with  no additional effort, we can make use of the connection between $E_{\rm I}$ and $E^{(2)}_{\rm dd}$ to write down
\beq\label{deltaEIWC}
\frac{\Delta \varepsilon_{\rm I}}{\alpha^2} = - 3 \tq e^{- \tq \tilde{\gamma}}\lambda_0^2~,
\eeq
which leads directly to
\bea\label{deltaEBL_3}
\frac{\Delta \varepsilon^{\rm TWC}_{\rm BL}}{\alpha^2} &=&
\frac{3}{2}\lambda_0^2 +\lambda_{\rm vW} \frac{3 \tq^2}{4} +
\left[ \frac{8}{\pi} - \frac{3}{2} \frac{\tq}{\lambda_0} \right]      \lambda_0^3
 -\frac{1}{4}\lambda_0^4 \left[ 9 \ln [f_0] + \frac{33}{8}\lambda_0 A + \frac{3}{8} \lambda_0^2 B\right]
 - 3 \tq e^{- \tq \tilde{\gamma}}\lambda_0^2~.
\eea
Armed with expressions for the shifts for both the 1DSP, Eq.~\eqref{deltaEBL_2}, and a TWC, Eq.~\eqref{deltaEBL_3}, 
we are now poised to investigate the phase diagram for the bilayer 2D dipolar Fermi gas.

\section{Phase Diagram of the bilayer system}\label{phase_diagram}

In contrast to the single-layer problem, there are now two independent variables which may be varied; namely, the coupling, $\lambda_0$, and the separation between the two layers, $\gamma$.
Consequently, our phase diagram will be an exploration of how the system responds to changes in these two variables.
As discussed in Ref.~\cite{vanzyl1_2015}, for the 1DSP, we take $\tilde{q} = 2 \lambda_0$, while for the
TWC we set $\tilde{q} = (8\pi/\sqrt{3})^{1/2}\lambda_0/2$.

\begin{figure}[ht]
\centering \scalebox{0.6}
{\includegraphics{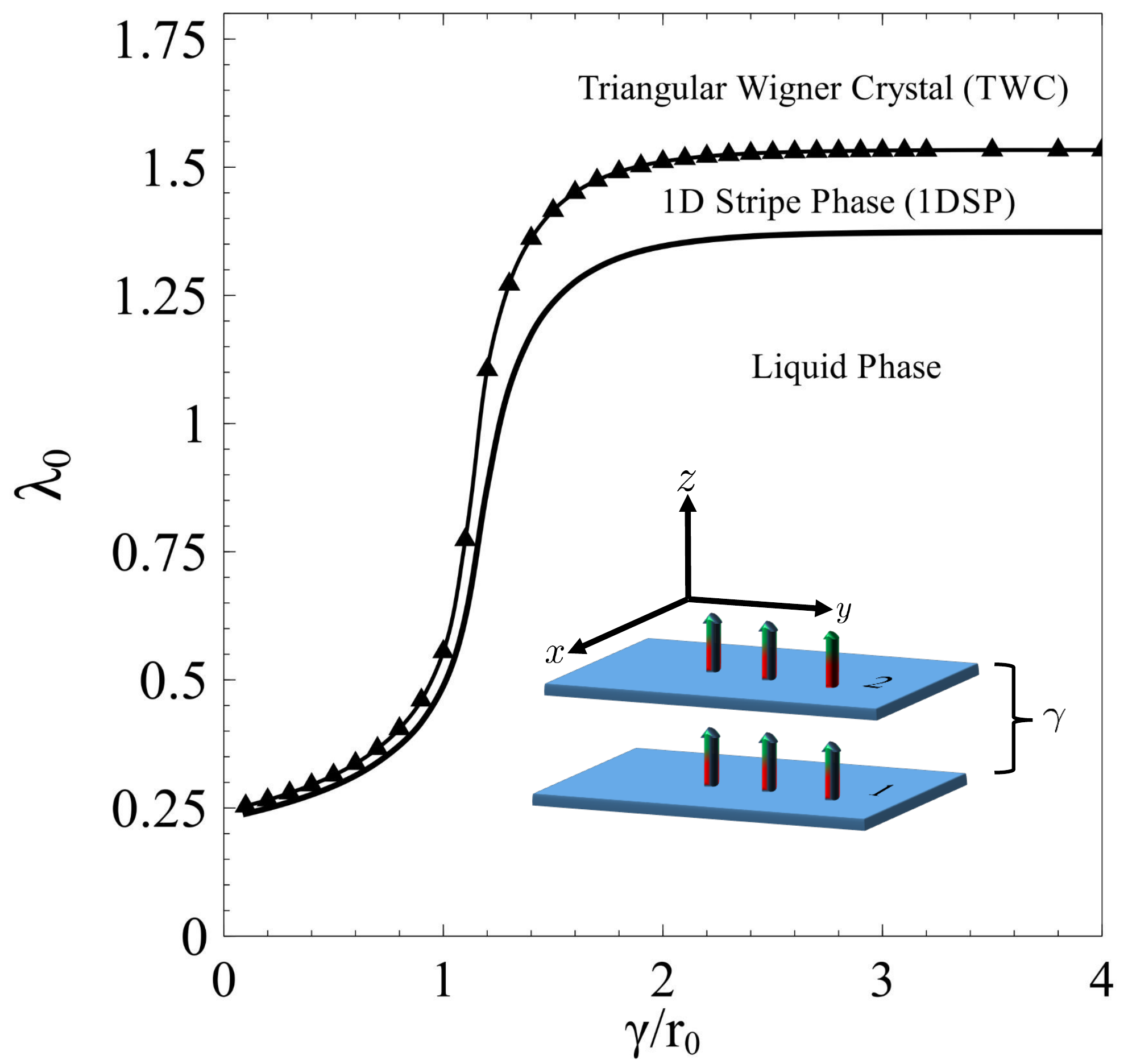}}
\caption{Phase diagram of the bilayer system.  As discussed in the text, $\lambda_0$ is the dimensionless
coupling and $r_0$ is the dipolar length scale.  For
$\gt = \gamma /r_0 \gtrsim 4$, the two layers are effectively uncoupled, and we reproduce the results found in Ref.~\cite{vanzyl1_2015}. The solid curve delineates the boundary between the liquid and 1DSP, while the solid curve with
triangles separates the 1DSP from the TWC.  Inset:  A schematic depiction of the model
under consideration.  In each 2D layer, the atoms have their dipole moments (represented by the arrows) 
aligned with the $z$-axis.  The layers are separated by a distance $\gamma$.}
\label{fig1}
\end{figure}

Specifically,
we look for a change in the sign of $\Delta \varepsilon^{\rm 1DSP}_{\rm BL}$ and 
$\Delta \varepsilon^{\rm TWC}_{\rm BL}$ as we vary $\lambda_0$ for a fixed layer separation, $\gamma$.  As it happens, $\Delta \varepsilon^{\rm 1DSP}_{\rm BL}$ {\em always}
crosses zero first (indicating that the 1DSP  phase is now energetically favourable over the uniform state) while 
$\Delta \varepsilon^{\rm TWC}_{\rm BL}$ remains positive.  Then, as $\lambda_0$ is further increased, 
$\Delta \varepsilon^{\rm TWC}_{\rm BL}$ also crosses zero (i.e., becomes negative), and we determine the preferred
inhomogeneous phase by comparing the values of $\Delta \varepsilon^{\rm 1DSP}_{\rm BL}$ and 
$\Delta \varepsilon^{\rm TWC}_{\rm BL}$; if $\Delta \varepsilon^{\rm TWC}_{\rm BL} < \Delta \varepsilon^{\rm 1DSP}_{\rm BL}$, the TWC phase is the ordered state, even though both phases are lower in energy than the uniform
Fermi liquid.  The procedure is repeated for different layer separations, leading to the phase diagram shown in
Figure~\ref{fig1}.

Let us now take a closer look at Fig.~\ref{fig1}.  First, 
we observe that for values of $\tilde{\gamma}=\gamma/r_0 \gtrsim 4$, the layers are effectively uncoupled, and
the results of Ref.~\cite{vanzyl1_2015} are recovered, viz., the system undergoes a transition first to a 1DSP followed
by a TWC at higher coupling.  Next, consider the case where we have a fixed layer separation, $\gt$, and start with
small $\lambda_0$, which increases as we move up vertically in the phase diagram.  It is evident that the window in which the 1DSP precedes the TWC gets smaller and smaller as the fixed layer separation decreases.
Indeed, for say $\gt =0.5$, the range of $\lambda_0$ distinguishing the 1DSP from the TWC is very narrow.  
This is perhaps
not so surprising given that the $\Delta \varepsilon_{\rm I}$ contribution has a maximum negative component as
$\gt \to 0$.

If we now fix the coupling, $\lambda_0$, and move horizontally in the phase diagram (i.e., varying $\gt$), we
see that there are layer separations at which a crystal or 1DSP is the preferred phase.
For example, if we take $\lambda_0=1$, and move horizontally in our phase diagram starting from small $\gt$,
we observe that up to $\gt \approx 1$, each layer possesses a TWC as the ground state of the system.  The system then
undergoes a transition to a 1DSP, which lasts over only a very short coupling range, 
followed by 
the system melting to a liquid.  Looking now at values of $1.4 \lesssim \lambda_0 \lesssim 1.5$, we see that the
liquid state is never present, and the system goes from a 1DSP to a TWC as the layer separation is decreased
(i.e., moving right-to-left in the phase diagram).
Finally, notice that for $\lambda_0 \gtrsim 1.5$, each layer remains frozen in a TWC,
regardless of the layer separation, $\gt$.

We have also confirmed that there are {\em no other} crystal symmetries more energetically favourable than the TWC, which
is in stark contrast to a bilayer 2D electron system~\cite{peeters}, where depending on the coupling and layer separation,
the TWC is {\em not} necessarily the ground state of the system.  It appears that, at least within our DFT, the only two
phases possessed by the bilayer 2D dFG are the 1DSP and the TWC. 

Our phase diagram also suggests that it should be easier to experimentally 
investigate density instabilities in bilayer configurations
owing to the fact that the transition from the Fermi liquid to the ordered state occurs at lower coupling as the spacing
between the layers, $\gt$, is decreased.  This is important from an experimental point of view, where the small 
$\lambda_0$ regime is much easier to achieve, owing to the rather small dipole moments of atoms and molecules
 currently employed in cold-atoms research.


\section{Conclusions and Closing Remarks}\label{closing}

We have extended the density-functional theory of Ref.~\cite{vanzyl1_2015} to the case of a bilayer, two-dimensional
dipolar Fermi gas. 
To our knowledge, the present density-functional study is the
 first of its kind to investigate
the possibility of a quantum phase transition to either a stripe, or Wigner crystal phase,  
in the bilayer two-dimensional dipolar Fermi gas.
While developing our formal extension of the work in Ref.~\cite{vanzyl1_2015}, we have found
an interesting connection between the direct
interlayer interaction energy, Eq.~\eqref{EIFT}, and the
{\em nonlocal} Hartree-Fock energy, Eq.~\eqref{hfE2};  namely, that the nonlocal Hartree-Fock energy can be obtained
by considering the direct interaction between two different layers, separated by a distance $\gamma$, and then
taking the $\gamma \to 0$ limit.   We have provided a physical interpretation of this connection, which is that  
$E^{(2)}_{\rm dd}$ represents the classical self-interaction energy 
in an inhomogeneous two-dimensional dipolar Fermi gas.
We have also presented analytical expressions, Eqs.~\eqref{deltaEBL_2} and~\eqref{deltaEBL_3}, which  allow for a numerically
efficient examination of the phase diagram for the bilayer system.
The relative ease of obtaining the phase diagram for the 
bilayer system should be viewed as a testament to the utility of the density-functional theory approach.

Our phase diagram (see Fig.~\ref{fig1}) demonstrates that the bilayer two-dimensonal dipolar Fermi gas has only two energetically 
favourable inhomogeneous phases, viz., a one-dimensional
stripe phase  or a triangular Wigner crystal in each layer.   In contrast to the case of the bilayer electron system
(see Ref.~\cite{peeters}), the stripe and Wigner crystal phases are perfectly commensurate in the two layers, with
no possibility of other crystalline symmetry phases existing owing to their high energy cost relative to the triangular
Wigner crystal.
Through an exploration of the $(\gt,\lambda_0)$-space, we have also confirmed that there are coupling regimes for 
which the system is always in a Wigner crystal or
stripe phase, completely bypassing the liquid state, regardless of the separation between the layers.  
We have also shown that the range of coupling, $\lambda_0$, over which the phase transition 
occurs can be significantly
reduced in a bilayer geometry, and have provided a simple explanation for this result in terms of the connection between
the nonlocal Hartree-Fock and interlayer interaction energy terms.

To close this paper, we mention that an interesting progression of this work would be to consider other exotic 
Fermi systems, such as the electron-hole
bilayer problem~\cite{elechole2,ye,elechole}, which to date has not been studied within a density-functional
theory approach.  It is also still an outstanding problem to provide a density-functional theory in the case where
the dipole moments of the atoms are canted at some angle relative to the $z$-axis (i.e., taking into account
the full anisotropic nature of the dipole-dipole interaction).
Finally, the development of our density-functional theory to finite-temperatures would be of interest, as
it is expected that at non-zero temperatures, the physics of the  Berezinskii-Kosterlitz-Thouless 
transition~\cite{bkt} is the appropriate description of the system.

\acknowledgements
This work was supported by grants from the Natural Sciences and Engineering Research Council of Canada (NSERC). W. Ferguson would like to thank the NSERC Undergraduate Student Research
Award (USRA) for additional financial  support.  BvZ would like to thank Prof.~Eugene Zaremba for
very useful conversations involving the nonlocal Hartree-Fock interaction energy.

\end{document}